\documentclass{amsart}%
\usepackage{amssymb}
\usepackage{amsfonts,amsmath}%
\usepackage{amsmath}%
\usepackage{amsfonts}%
\usepackage{geometry}
\usepackage{graphicx}
\providecommand{\U}[1]{\protect \rule{.1in}{.1in}}
\newtheorem{theorem}{Theorem}
\theoremstyle{plain}

\newtheorem{definition}{Definition}
\newtheorem{example}{Example}

\newtheorem{proposition}{Proposition}
\newtheorem{remark}{Remark}

\numberwithin{equation}{section}
\begin{document}
\title{New generalizations of circular complex fuzzy sets and Gaussian weighted aggregation operators}
\author{Yelda G\"{U}LFIRAT}
\curraddr{Graduate School of Natural and Applied Sciences, Ankara University, Ankara, T\"{u}rkiye.}
\email{yeldaratt@gmail.com}
\author{Mehmet \"{U}NVER}
\curraddr{Department of Mathematics, Faculty of Science, Ankara University, Ankara, T\"{u}rkiye.}
\email{munver@ankara.edu.tr}


\begin{abstract}
In this paper, we introduce the concept of the circular complex $q$-rung orthopair fuzzy set (CC$q$-ROFS) as a novel generalization that unifies the existing frameworks of circular complex intuitionistic fuzzy sets (CCIFSs) and complex $q$-rung orthopair fuzzy sets.
If $q = 2$, the structure is referred to as a circular complex Pythagorean fuzzy set, and if $q = 3$, it is called a circular complex Fermatean fuzzy set.
The proposed approach extends the Gaussian-based framework to the CC$q$-ROFSs, aiming to achieve a smoother and statistically meaningful representation of uncertainty. Within this setting, new Gaussian-based aggregation operators for CC$q$-ROFSs are constructed by employing the Gaussian triangular norm and conorm. Furthermore, Gaussian-weighted arithmetic and Gaussian-weighted geometric aggregation operators are formulated to enable consistent integration of membership and non-membership information for fuzzy modeling and decision-making.

\end{abstract}
\maketitle

\section{Introduction}

Since its introduction by Zadeh \cite{zadeh}, fuzzy set theory has served as
a fundamental mathematical tool for representing uncertainty, imprecision, and
vagueness inherent in real-world problems. Over the decades, it has evolved
from a theoretical construct into a versatile framework extensively applied in
decision-making, control systems, pattern recognition, and artificial
intelligence. While classical fuzzy sets consider only the degree of
membership, Atanassov \cite{atanassov 86} extended this foundation by
introducing the concept of intuitionistic fuzzy sets (IFSs), where both
membership and non-membership degrees coexist. This dual representation
provides additional flexibility in modeling hesitation and partial truth,
thereby establishing IFSs as one of the most significant generalizations in
fuzzy set theory.

Later, Yager \cite{yager 2016} proposed the $q-$rung orthopair fuzzy sets
($q$-ROFSs), a unified framework that encompasses both intuitionistic fuzzy
sets and Pythagorean fuzzy sets (PFSs) \cite{yager2013a,yager2013b} and Fermatean fuzzy sets \cite{fermatean}. This development significantly improved the expressive power
of fuzzy modeling, enabling more flexible and adaptive control over the
representation of uncertainty. Collectively, these advances have strengthened
the theoretical foundation and computational robustness of fuzzy logic,
particularly in multi-criteria decision-making (MCDM) and classification problems.

A further conceptual leap occurred with the introduction of complex fuzzy sets by Ramot et al. \cite{ramot2002}, who extended membership degrees to
the complex plane. By incorporating both magnitude and phase components, this
model captured dimensions of uncertainty that classical real-valued fuzzy sets
could not represent---an advantage particularly beneficial in applications
such as signal processing, time-series analysis, and dynamic system modeling.
Subsequent research merged the intuitionistic framework with complex
representation, giving rise to complex intuitionistic fuzzy sets (see,
\cite{alkouri}), which are capable of simultaneously modeling hesitation and phase-dependent ambiguity. Moreover, Ullah et al. \cite{ullah} pioneered the concept of complex PFS, investigating their distance measures and applying them in pattern recognition. The concept of complex $q$-rung orthopair fuzzy set, which generalizes the CPFS, was utilized by Ali and Mahmood \cite{ali} to develop Maclaurin symmetric mean operators and apply them to environmental problems.

To improve the geometric flexibility of representing fuzzy membership and non-membership information, Atanassov \cite{atanassov20} introduced the concept of circular intuitionistic fuzzy set (CIFS), in which the membership degrees are represented on a circle rather than by a single scalar value. This circular formulation enables uncertainty to be characterized as a region instead of a point estimate, analogous to a confidence interval in statistics, thereby providing a more comprehensive and intuitive depiction of imprecise information. Motivated by this idea, Bozyiğit et al. \cite{bozyigit} proposed the circular PFS, while Yusoof et al. \cite{yusoof} further extended the framework by introducing the concept of the circular $q$-rung orthopair fuzzy set.

Building upon these foundations, Garg et al. \cite{garg25} unified the complex and circular paradigms to propose the circular complex intuitionistic fuzzy set (CCIFS). This hybrid structure integrates magnitude--phase representation with circular flexibility, offering a
multidimensional and geometrically interpretable framework of uncertainty
modeling. Despite its expressive power, the CCIFS framework remains restricted by the intuitionistic constraint on the real and imaginary parts of the membership degrees, which limits its adaptability in problems demanding higher tolerance for hesitation and greater control over the geometric distribution of uncertainty.
To address this limitation and further develop representational richness, we extend the circular complex framework to the $q$-rung orthopair fuzzy domain, introducing the circular complex $q$-rung orthopair fuzzy sets (CC$q$-ROFS). As special cases, we define circular complex Pythagorean fuzzy sets (CCPFSs) and circular complex Fermatean fuzzy sets (CCFFSs). These generalizations not only preserve the geometric intuition of circular and complex formulations but also incorporate the flexibility of the $q$-rung exponent, allowing a broader range of membership–non-membership configurations and offering a more refined balance between precision and hesitancy in decision-making and information modeling.

At the core of fuzzy reasoning, triangular norms (t-norms) and triangular conorms (t-conorms) play a crucial role
in aggregation and inference processes. Classical operators, such as minimum,
product, Hamacher, Einstein, Dombi, and Lukasiewicz (see,
\cite{dombi,hamacher,klment2004a,klement2004}) have long been used to model conjunction and
disjunction in fuzzy logic. In this context, several extensions have emerged. Bobillo and Straccia
\cite{bobillo} applied product t-norms within fuzzy description logic,
Dimuro et al. \cite{dimuro} developed the CMin-integral, based on the
minimum t-norm, for fuzzy rule-based classification, and Garg \cite{garg17} introduced generalized Pythagorean fuzzy geometric operators derived from
the Einstein family. Recently, Ünver \cite{unver25} introduced the concept of Gaussian t-norms and t-conorms and their applications to intuitionistic fuzzy classification, presenting a novel approach for fuzzy information fusion. This development highlighted the usefulness of Gaussian-based operators owing to their smoothness, differentiability, and statistical interpretability. In particular, the Gaussian error function has been utilized to construct Archimedean t-norms and t-conorms, enabling continuous and probabilistically meaningful aggregation within fuzzy systems.

Motivated by these advancements, the present study extends the Gaussian-based framework of \"{U}nver \cite{unver25} into the domain of CC$q$-ROFSs. The integration of the Gaussian error function with CC$q$-ROFS combines statistical smoothness with
multidimensional fuzzy expressivity, yielding a novel and powerful approach for modeling uncertainty, aggregation, and classification in complex decision environments. This work not only generalizes existing models but also bridges
theoretical development with practical applicability in advanced fuzzy decision-making systems.

The remainder of the paper is organized as follows: Section \ref{sec2} recalls the basic notions used throughout the study. In Section \ref{sec3}, the algebraic operations of the CC$q$-ROFS are first defined, and then the aggregation operators are constructed using Gaussian t-norm and t-conorm.

\section{Preliminaries}\label{sec2}

In this section, we recall some fundamental concepts and notations that will
be used throughout the paper. The definitions presented in this section
provide the mathematical foundation for constructing the proposed $q-$rung
circular complex intuitionistic fuzzy framework where $q\geq1$. Furthermore
the concepts directly related to intuitionistic fuzzy, $q-$rung orthopair and
circular complex fuzzy structures are introduced, as they form both the
motivation and the groundwork for the theoretical developments presented in
the subsequent sections.

Throughout this paper, $X=\{x_{1},...,x_{n}\}$ denotes a finite set and the
unit interval $[0,1]$ is represented by $\mathcal{I}$. We start with the definition of several fuzzy sets that we motivated.

\begin{definition}
\cite{atanassov 86}\label{defatanassov} An IFS\ $A$ on $X$ is defined by a
membership function $\mu_{A}:X\rightarrow \mathcal{I}$ and a non-membership
function $\nu_{A}:X\rightarrow \mathcal{I}$, which satisfy
\[
0\leq \mu_{A}\left(  x_{i}\right)  +\nu_{A}\left(  x_{i}\right)  \leq1,
\]
for any $i=1,...,n$, where $\mathcal{I}=[0,1]$.

\end{definition}

\begin{definition}\label{q}
\cite{yager 2016} Let $q\ge 1$ be a real number. A $q$-ROFS $A$ on $X$ is
defined by a membership function $\mu_{A}:X\rightarrow$ $\mathcal{I}$ and a
non-membership function $\nu_{A}:X\rightarrow \mathcal{I}$, which satisfy
\[
0\leq \left(  \mu_{A}\left(  x_{i}\right)  \right)  ^{q}+\left(  \nu_{A}\left(
x_{i}\right)  \right)  ^{q}\leq1,
\]
$\ $ for any $i=1,...,n$.

\end{definition}

When $q=1$, the structure given in Definition \ref{q} reduces to the IFS
\cite{atanassov 86}; for $q=2$ it
corresponds to the PFS \cite{yager2013a, yager2013b}; for $q=3$ it reduces to Fermatean fuzzy set \cite{fermatean}; and for higher values of $q$ it allows greater flexibility
in modeling uncertainty by enlarging the admissible domain of membership and
non-membership degrees.

\begin{definition}
\cite{atanassov20} Let $r\in \mathcal{I}$. A CIFS $A_{r}$ in$\ X$ is defined by
\[
A_{r}=\left \{  \left \langle x_{i},\  \mu_{A}\left(  x_{i}\right)  ,\  \nu
_{A}\left(  x_{i}\right)  ;r\right \rangle :i=1,...,n\right \}  ,
\]
where $\mu_{A},\nu_{A}:X\rightarrow \mathcal{I}$ are functions such that
\[
0\leq \mu_{A}\left(  x_{i}\right)  +\nu_{A}\left(  x_{i}\right)  \leq1.
\]
$r$ is the radius of the circle around the point on the plane. This circle
represents the membership and non-membership degree of $x\in X.$
\end{definition}

\begin{definition}
\cite{garg25} A CCIFS $A$ on $X$ is defined as
\[
A=\left \{  \left \langle x_{i},\  \mu_{A}\left(  x_{i}\right)  ,\  \nu_{A}\left(
x_{i}\right)  ,r_{A}\left(  x_{i}\right)  \right \rangle :i=1,...,n\right \}
\]
where,
\begin{align*}
\mu_{A}\left(  x_{i}\right)   & =\left(  \mu_{A}^{\operatorname{Re}}\left(
x_{i}\right)  ,\mu_{A}^{\operatorname{Im}}\left(  x_{i}\right)  \right)  \in \mathcal{I}^2,\  \\
\nu_{A}\left(  x_{i}\right)   & =\left(  \nu_{A}^{\operatorname{Re}}\left(
x_{i}\right)  ,\nu_{A}^{\operatorname{Im}}\left(  x_{i}\right)  \right)  \in \mathcal{I}^2,
\end{align*}
are complex numbers satisfying
\begin{align*}
0  & \leq \mu_{A}^{\operatorname{Re}}(x)+\nu_{A}^{\operatorname{Re}}%
(x)\leq1,\medskip \\
0  & \leq \mu_{A}^{\operatorname{Im}}(x)+\nu_{A}^{\operatorname{Im}}%
(x)\leq1.\medskip
\end{align*} and the complex number
\[
r_{A}\left(  x_{i}\right)  =(r_{A}^{\operatorname{Re}}\left(  x_{i}\right)
,r_{A}^{\operatorname{Im}}\left(  x_{i}\right)  )\in \mathcal{I}^2
\]
denotes the radius of the circle corresponding to the element $x_{i}\in X$ in
the complex plane. Here $\mu_{A}$ and $\nu_{A}$ are called the membership and non-membership
functions, respectively.
\end{definition}

Let us recall the following three concepts, which are used in defining
algebraic operations.

\begin{definition}
\cite{klement05, schweizer} A t-norm is a function $\tau
:\mathcal{I}^{2}\rightarrow \mathcal{I}$ that satisfies \newline(1)
$\tau \left(  x,1\right)  =x$ for any $x\in \mathcal{I}$ (border
condition),\newline(2) $\tau \left(  x,y\right)  =\tau \left(  y,x\right)  $ for
any $x,\ y\in \mathcal{I}$ (commutativity),\newline(3) $\tau \left(
x,\tau(y,z)\right)  =\tau \left(  \tau(x,y),z\right)  $ for any $x,\ y,\ z\in
\mathcal{I}$ (associativity),\newline(4) $\tau \left(  x,y\right)  \leq$
$\tau \left(  x^{\prime},y^{\prime}\right)  $ whenever $x\leq x^{\prime}$ and
$y\leq y^{\prime}$ for any $x,\ x^{\prime},\ y,\ y^{\prime}\in \mathcal{I}$ (monotonicity).
\end{definition}

\begin{definition}
\cite{klement05, schweizer} A t-conorm is a function $\rho
:\mathcal{I}^{2}\rightarrow \mathcal{I}$ that satisfies\newline(1) $\rho \left(
x,0\right)  =x$ for any $x\in \mathcal{I}$ (border condition),\newline(2)
$\rho \left(  x,y\right)  =\rho \left(  y,x\right)  $ for any $x,\ y\in
\mathcal{I}$ (commutativity),\newline(3) $\rho \left(  x,\tau(y,z)\right)
=\rho \left(  \tau(x,y),z\right)  $ for any $x,\ y,\ z\in \mathcal{I}$
(associativity),\newline(4) $\rho \left(  x,y\right)  \leq$ $\rho \left(
x^{\prime},y^{\prime}\right)  $ whenever $x\leq x^{\prime}$ and $y\leq
y^{\prime}$ for any $x,\ x^{\prime},\ y,\ y^{\prime}\in \mathcal{I}$ (monotonicity).
\end{definition}

A continuous t-norm $\tau$ is defined as Archimedean if it satisfies the
condition $\tau \left(  x,x\right)  \leq x$ for all $x\in \mathcal{I}$. This
property ensures that such t-norms remain compatible with various operations
employed within the framework of fuzzy logic. In particular, the fuzzy
implication and negation operations derived from these t-norms preserve
fundamental characteristics such as monotonicity and involutiveness, which are
essential in many fuzzy inference systems.

To obtain certain specific types of t-norms, the concept of an additive
generator can be utilized. This approach allows a t-norm to be expressed
through a strictly decreasing function, thereby enabling the systematic
construction of new families of t-norms.

\begin{definition}
(\cite{klement2002},\cite{klement2004}) \label{klement34}A strictly decreasing
function $g:\mathcal{I}\rightarrow \left[  0,\infty \right]  $ with $g\left(
1\right)  =0$ is called the additive generator of a t-norm $\tau$ if we have
$\tau \left(  x,y\right)  =g^{-1}\left(  g\left(  x\right)  ,g\left(  y\right)
\right)  $ for all $\left(  x,y\right)  \in \mathcal{I}^{2}.$
\end{definition}

\begin{theorem}
(\cite{klement2004}) Let $\tau$ be a t-norm on $\mathcal{I}$. The following
statements are equivalent:\newline(i) $\tau$ is a continuous Archimedean
t-norm.\newline(ii) $\tau$ has a continuous additive generator, i.e. there
is a continuous, strictly decreasing function $g:\mathcal{I}\rightarrow \left[
0,\infty \right]  $ with $\tau \left(  1\right)  =0$ such that $\tau \left(
x,y\right)  =g^{-1}\left(  g\left(  x\right)  +g\left(  y\right)  \right)  $
for all $\left(  x,y\right)  \in \mathcal{I}^{2}.$
\end{theorem}

\begin{remark}
(\cite{klement2004}) The dual t-conorm $\rho$ associated with a continuous
Archimedean t-norm $\tau$ also qualifies as a continuous Archimedean
t-conorm, for all $x\in(0,1),$ $\rho(x,x)>x.$ The t-conorm $\rho$ is
characterized by an additive generator $h:\mathcal{I}\rightarrow \left[
0,\infty \right]  $ defined by $h(x)=g(1-x)$ which effectively mirrors the
generator $g$ used in the t-norm. Using this generator, the t-conorm is
given by
\[
\rho(x,y)=h^{-1}(h(x)+h(y)).
\]

\end{remark}

\section{Circular complex $q$-rung fuzzy sets and aggregation
operators}\label{sec3}

This section is devoted to the extension of the CCIFSs, introduced by Garg et al. in
\cite{garg25}, to a more general framework of CC$q$-ROFSs where $q\geq1$. Notably, when $q=1$,
the proposed structure reduces to one type of the original formulation
introduced by them.

Furthermore, in developing the algebraic structure and aggregation operators
within the proposed framework, we employ the Gaussian Archimedean t-norm and
t-conorm very recently introduced by \cite{unver25}. These Gaussian-based
operations provide a smooth, flexible, and analytically tractable foundation
for defining the corresponding fuzzy aggregation operators in the CC$q$-ROFS environment.

\begin{definition}
A $q-$ CCIFS $A$ in $X$ is defined by
\[
A=\{(\mu_{A}\left(  x\right)  ,\nu_{A}\left(  x\right)  ,r_{A}\left(
x\right)  ):x\in X\},
\]
where
\begin{align*}
\mu_{A}(x)  & =(\mu_{A}^{\operatorname{Re}}(x),\mu_{A}^{\operatorname{Im}%
}(x)),\medskip \\
\nu_{A}(x)  & =(\nu_{A}^{\operatorname{Re}}(x),\nu_{A}^{\operatorname{Im}%
}(x)),\medskip \\
r_{A}(x)  & \in[0,1],\medskip
\end{align*}
and $\mu_{A}^{\operatorname{Re}},\mu_{A}^{\operatorname{Im}},\nu
_{A}^{\operatorname{Re}},\nu_{A}^{\operatorname{Im}}:X\rightarrow \lbrack0,1]$
are functions satisfying the following conditions%
\begin{align*}
0  & \leq \left[  \mu_{A}^{\operatorname{Re}}(x)\right]  ^{q}+\left[  \nu
_{A}^{\operatorname{Re}}(x)\right]  ^{q}\leq1,\medskip \\
0  & \leq \left[  \mu_{A}^{\operatorname{Im}}(x)\right]  ^{q}+\left[  \nu
_{A}^{\operatorname{Im}}(x)\right]  ^{q}\leq1.\medskip
\end{align*}
Here, $\mu_{A},\nu_{A}$ and $r_{A}$ are called the membership function and the
non-membership function and the radius around the circle of membership degrees, respectively. Additionally, the neutral function is defined by
\[
\pi(x)=(\pi^{\operatorname{Re}}(x),\pi^{\operatorname{Im}}(x))\text{,}%
\]
where
\begin{align*}
\pi^{\operatorname{Re}}(x)  & =\left[  1-\left(  \mu_{A}^{\operatorname{Re}%
}(x)\right)  ^{q}-(\nu_{A}^{\operatorname{Re}}(x))^{q}\right]  ^{\frac{1}{q}%
},\medskip \\
\pi^{\operatorname{Im}}(x)  & =\left[  1-\left(  \mu_{A}^{\operatorname{Im}%
}(x)\right)  ^{q}-(\nu_{A}^{\operatorname{Im}}(x))^{q}\right]  ^{\frac{1}{q}%
},\medskip
\end{align*}
and for $i=1,...,n$, the triplet
\[
A_{i}=(\mu_{A_{i}}\left(  x\right)  =(\mu_{A_{i}}^{\operatorname{Re}}%
(x),\mu_{A_{i}}^{\operatorname{Im}}(x)),\nu_{A_{i}}\left(  x\right)
=(\nu_{A_{i}}^{\operatorname{Re}}(x),\nu_{A_{i}}^{\operatorname{Im}%
}(x)),r_{A_{i}}\left(  x\right)  )
\]
is called a circular complex $q$-rung orthopair fuzzy value (CC$q$-ROFV).
\end{definition}
We state that the new concept of CC$q$-ROFS
extends the concept of CCIFS recently studied in \cite{garg25}. If $q = 2$, the structure is referred to as a CCPFS instead of a CC$2$-ROFS, and if $q = 3$, it is called a CCFFS instead of a CC$3$-ROFS.

\begin{example}
Let $X=\left \{  x_{1},x_{2},x_{3}\right \}  $ be a finite set and consider
$q=2.$ A CCPFS on $X$ is defined by
\[%
\begin{array}
[c]{lll}%
A & = & \{ \left \langle x_{1},(0.5,0.3),(0.4,0.2),0.2\right \rangle ,\\
&  & \  \left \langle x_{2},(0.6,0.2)),(0.4,0.2),0.3\right \rangle ,\\
&  & \  \left \langle x_{3},(0.4,0.4)),(0.5,0.3),0.1\right \rangle \} \text{.}%
\end{array}
\]
For each element $i=1,2,3,$ the $q-$rung conditions are satisfied for $q=2$ as
\[%
\begin{tabular}
[c]{c}%
$0\leq \mu_{A}^{\operatorname{Re}}(x_{i})^{2}+\nu_{A}^{\operatorname{Re}}%
(x_{i})^{2}\leq1,\medskip$\\
$0\leq \mu_{A}^{\operatorname{Im}}(x_{i})^{2}+\nu_{A}^{\operatorname{Im}}%
(x_{i})^{2}\leq1,\medskip$\\
$r_{A}(x_{i})\in \lbrack0,1].$%
\end{tabular}
\
\]

\end{example}

Now, we define a set of algebraic operations for CC$q$-ROFVs by utilizing the Gaussian-based Archimedean t-norm and t-conorm as in \cite{unver25}. These operations establish the mathematical foundation for the development of
the proposed Gaussian aggregation operators,
thereby enabling smooth and consistent modeling of uncertainty in
complex-valued fuzzy environments. Motivated by \cite{unver25}, we now introduce algebraic operations between CC$q$-ROFVs.

\begin{definition}
\label{def16}Let $A_{1},$ $A_{2}$ be two CC$q$-ROFVs, let $\lambda \geq0$ be
a real number, and let $Z:\left[  0,1\right]  \rightarrow \left[  0,\infty \right]  $ be the additive generator of a continuous Archimedean t-norm or t-conorm. Then we define \newline i) the summation by
\[%
\begin{array}
[c]{lll}%
A_{1}\oplus_{q}A_{2} & = & \left \langle \left(  \left[  1-\left(
\operatorname{erf}^{-1}\left(  \frac{\operatorname{erf}\left(  1-\left(
\mu_{A_{1}}^{\operatorname{Re}}\right)  ^{q}\right)  \operatorname{erf}\left(
1-\left(  \mu_{A_{2}}^{\operatorname{Re}}\right)  ^{q}\right)  }%
{\operatorname{erf}\left(  1\right)  }\right)  \right)  ^{q}\right]
^{^{\frac{1}{q}}}\right.  \right.  ,\\
&  & \  \  \  \  \  \  \  \  \left.  \left[  1-\left(  \operatorname{erf}^{-1}\left(
\frac{\operatorname{erf}\left(  1-\left(  \mu_{A_{1}}^{\operatorname{Im}%
}\right)  ^{q}\right)  \operatorname{erf}\left(  1-\left(  \mu_{A_{2}%
}^{\operatorname{Im}}\right)  ^{q}\right)  }{\operatorname{erf}\left(
1\right)  }\right)  \right)  ^{q}\right]  ^{\frac{1}{q}}\right)  ,\\
&  & \left(  \left[  \operatorname{erf}^{-1}\left(  \frac{\operatorname{erf}%
\left(  \nu_{A_{1}}^{\operatorname{Re}}\right)  ^{q}\operatorname{erf}\left(
\nu_{A_{2}}^{\operatorname{Re}}\right)  ^{q}}{\operatorname{erf}\left(
1\right)  }\right)  \right]  ^{\frac{1}{q}},\left[  \operatorname{erf}%
^{-1}\left(  \frac{\operatorname{erf}\left(  \nu_{A_{1}}^{\operatorname{Im}%
}\right)  ^{q}\operatorname{erf}\left(  \nu_{A_{2}}^{\operatorname{Im}%
}\right)  ^{q}}{\operatorname{erf}\left(  1\right)  }\right)  \right]
^{\frac{1}{q}}\right)  ,\medskip \\
&  & \left.  Z\left(  r_{1},r_{2}\right)
\begin{tabular}
[c]{l}%
\\
\\
\end{tabular}
\right \rangle ,
\end{array}
\]
where $Z\left(  r_{1},r_{2}\right)  =Z^{-1}\left(  Z\left(  r_{1}\right)
+Z\left(  r_{2}\right)  \right)  .\medskip$\newline ii) the multiplication by
\[%
\begin{array}
[c]{lll}%
A_{1}\otimes_{q}A_{2} & = & \left \langle \left(  \left[  \operatorname{erf}%
^{-1}\left(  \frac{\operatorname{erf}\left(  \mu_{A_{1}}^{\operatorname{Re}%
}\right)  ^{q}\operatorname{erf}\left(  \mu_{A_{2}}^{\operatorname{Re}%
}\right)  ^{q}}{\operatorname{erf}\left(  1\right)  }\right)  \right]
^{\frac{1}{q}},\left[  \operatorname{erf}^{-1}\left(  \frac{\operatorname{erf}%
\left(  \mu_{A_{1}}^{\operatorname{Im}}\right)  ^{q}\operatorname{erf}\left(
\mu_{A_{2}}^{\operatorname{Im}}\right)  ^{q}}{\operatorname{erf}\left(
1\right)  }\right)  \right]  ^{\frac{1}{q}}\right)  \right.  ,\medskip \\
&  & \left(  \left[  1-\left(  \operatorname{erf}^{-1}\left(  \frac
{\operatorname{erf}\left(  1-\left(  \nu_{A_{1}}^{\operatorname{Re}}\right)
^{q}\right)  \operatorname{erf}\left(  1-\left(  \nu_{A_{2}}%
^{\operatorname{Re}}\right)  ^{q}\right)  }{\operatorname{erf}\left(
1\right)  }\right)  \right)  ^{q}\right]  ^{\frac{1}{q}}\right.  ,\medskip \\
&  & \  \  \  \  \  \  \  \  \left.  \left[  1-\left(  \operatorname{erf}^{-1}\left(
\frac{\operatorname{erf}\left(  1-\left(  \nu_{A_{1}}^{\operatorname{Im}%
}\right)  ^{q}\right)  \operatorname{erf}\left(  1-\left(  \nu_{A_{2}%
}^{\operatorname{Im}}\right)  ^{q}\right)  }{\operatorname{erf}\left(
1\right)  }\right)  \right)  ^{q}\right]  ^{\frac{1}{q}}\right)  ,\\
&  & Z\left(  r_{1},r_{2}\right)  \left.
\begin{tabular}
[c]{l}%
\\
\\
\end{tabular}
\right \rangle .\medskip
\end{array}
\]
iii) the multiplication with a scalar by
\[%
\begin{array}
[c]{lll}%
\lambda_{q}A_{1} & = & \left \langle \left(  \left[  1-\operatorname{erf}%
^{-1}\left(  \frac{\left(  \operatorname{erf}\left(  1-\mu_{A_{1}%
}^{\operatorname{Re}}\right)  ^{q}\right)  ^{\lambda}}{\operatorname{erf}%
\left(  1\right)  ^{\lambda-1}}\right)  \right]  ^{^{\frac{1}{q}}},\left[
1-\operatorname{erf}^{-1}\left(  \frac{\left(  \operatorname{erf}\left(
1-\mu_{A_{1}}^{\operatorname{Im}}\right)  ^{q}\right)  ^{\lambda}%
}{\operatorname{erf}\left(  1\right)  ^{\lambda-1}}\right)  \right]
^{^{\frac{1}{q}}}\right)  ,\right.  \medskip \\
&  & \left(  \left[  \operatorname{erf}^{-1}\left(  \frac{\left(
\operatorname{erf}\left(  \nu_{A_{1}}^{\operatorname{Re}}\right)  ^{q}\right)
^{\lambda}}{\operatorname{erf}\left(  1\right)  ^{\lambda-1}}\right)  \right]
^{^{\frac{1}{q}}},\left[  \operatorname{erf}^{-1}\left(  \frac{\left(
\operatorname{erf}\left(  \nu_{A_{1}}^{\operatorname{Im}}\right)  ^{q}\right)
^{\lambda}}{\operatorname{erf}\left(  1\right)  ^{\lambda-1}}\right)  \right]
^{^{\frac{1}{q}}}\right)  ,\medskip \\
&  & Z\left(  r_{1},r_{2}\right)  \left.
\begin{array}
[c]{c}%
\\
\\
\end{array}
\right \rangle .\medskip
\end{array}
\]
\newline iv) the power by
\[%
\begin{array}
[c]{lll}%
A_{1}^{\lambda_{q}} & = & \left \langle \left(  \left[  \operatorname{erf}%
^{-1}\left(  \frac{\left(  \operatorname{erf}\left(  \mu_{A_{1}}%
^{\operatorname{Re}}\right)  ^{q}\right)  ^{\lambda}}{\operatorname{erf}%
\left(  1\right)  ^{\lambda-1}}\right)  \right]  ^{^{\frac{1}{q}}},\left[
\operatorname{erf}^{-1}\left(  \frac{\left(  \operatorname{erf}\left(
\mu_{A_{1}}^{\operatorname{Im}}\right)  ^{q}\right)  ^{\lambda}}%
{\operatorname{erf}\left(  1\right)  ^{\lambda-1}}\right)  \right]
^{^{\frac{1}{q}}}\right)  ,\right.  \medskip \\
&  & \left(  \left[  1-\operatorname{erf}^{-1}\left(  \frac{\left(
\operatorname{erf}\left(  1-\nu_{A_{1}}^{\operatorname{Re}}\right)
^{q}\right)  ^{\lambda}}{\operatorname{erf}\left(  1\right)  ^{\lambda-1}%
}\right)  \right]  ^{^{\frac{1}{q}}},\left[  1-\operatorname{erf}^{-1}\left(
\frac{\left(  \operatorname{erf}\left(  1-\nu_{A_{1}}^{\operatorname{Im}%
}\right)  ^{q}\right)  ^{\lambda}}{\operatorname{erf}\left(  1\right)
^{\lambda-1}}\right)  \right]  ^{^{\frac{1}{q}}}\right)  ,\medskip \\
&  & Z\left(  r_{1},r_{2}\right)  \left.  {}\right \rangle .\medskip
\end{array}
\]

\end{definition}

\begin{remark}
\label{remark17}In Definition \ref{def16}, considering the Gaussian t-norm and t-conorm defined in \cite{unver25} we use the properties of operations
involving CC$q$-ROFV as$\newline$
\[%
\begin{array}
[c]{lll}%
A_{1}\oplus_{q}A_{2} & = & \left \langle \left(  h^{-1}\left(  h\left(
\mu_{A_{1}}^{\operatorname{Re}}\right)  +h\left(  \mu_{A_{2}}%
^{\operatorname{Re}}\right)  \right)  ,h^{-1}\left(  h\left(  \mu_{A_{1}%
}^{\operatorname{Im}}\right)  +h\left(  \mu_{A_{2}}^{\operatorname{Im}%
}\right)  \right)  \right)  \right.  ,\medskip \\
&  & \left(  g^{-1}\left(  g\left(  \nu_{A_{1}}^{\operatorname{Re}}\right)
+g\left(  \nu_{A_{2}}^{\operatorname{Re}}\right)  \right)  ,g^{-1}\left(
g\left(  \nu_{A_{1}}^{\operatorname{Im}}\right)  +g\left(  \nu_{A_{2}%
}^{\operatorname{Im}}\right)  \right)  \right)  ,\medskip \\
&  & \left.  Z^{-1}(Z\left(  r_{1})+Z(r_{2}\right)  )\right \rangle ,\medskip
\end{array}
\]
$\bigskip \newline$%
\[%
\begin{array}
[c]{lll}%
A_{1}\otimes_{q}A_{2} & = & \left \langle \left(  g^{-1}\left(  g\left(
\mu_{A_{1}}^{\operatorname{Re}}\right)  +g\left(  \mu_{A_{2}}%
^{\operatorname{Re}}\right)  \right)  ,g^{-1}\left(  g\left(  \mu_{A_{1}%
}^{\operatorname{Im}}\right)  +g\left(  \mu_{A_{2}}^{\operatorname{Im}%
}\right)  \right)  \right)  \right.  ,\medskip \\
&  & \left(  h^{-1}\left(  h\left(  \nu_{A_{1}}^{\operatorname{Re}}\right)
+h\left(  \nu_{A_{2}}^{\operatorname{Re}}\right)  \right)  ,h^{-1}\left(
h\left(  \nu_{A_{1}}^{\operatorname{Im}}\right)  +h\left(  \nu_{A_{2}%
}^{\operatorname{Im}}\right)  \right)  \right)  ,\medskip \\
&  & \left.  Z^{-1}(Z\left(  r_{1})+Z(r_{2}\right)  )\right \rangle ,\medskip
\end{array}
\]
$\bigskip \newline$%
\[%
\begin{array}
[c]{lll}%
\lambda_{q}A_{1} & = & \left \langle \left(  h^{-1}\left(  \lambda h\left(
\mu_{A_{1}}^{\operatorname{Re}}\right)  \right)  ,h^{-1}\left(  \lambda
h\left(  \mu_{A_{1}}^{\operatorname{Im}}\right)  \right)  \right)  \right.
,\medskip \\
&  & \left(  g^{-1}\left(  \lambda g\left(  \nu_{A_{1}}^{\operatorname{Re}%
}\right)  \right)  ,g^{-1}\left(  \lambda g\left(  \nu_{A_{1}}%
^{\operatorname{Im}}\right)  \right)  \right)  ,\medskip \\
&  & \left.  Z^{-1}(Z\left(  r_{1})+Z(r_{2}\right)  )\right \rangle ,\medskip
\end{array}
\]
$\bigskip \newline$and%
\[%
\begin{array}
[c]{lll}%
A_{1}^{\lambda_{q}} & = & \left \langle \left(  g^{-1}\left(  \lambda g\left(
\mu_{A_{1}}^{\operatorname{Re}}\right)  \right)  ,g^{-1}\left(  \lambda
g\left(  \mu_{A_{1}}^{\operatorname{Im}}\right)  \right)  \right)  \right.
,\medskip \\
&  & \left(  h^{-1}\left(  \lambda h\left(  \nu_{A_{1}}^{\operatorname{Re}%
}\right)  \right)  ,h^{-1}\left(  \lambda h\left(  \nu_{A_{1}}%
^{\operatorname{Im}}\right)  \right)  \right)  ,\medskip \\
&  & \left.  Z^{-1}(Z\left(  r_{1})+Z(r_{2}\right)  )\right \rangle .\medskip
\end{array}
\]

\end{remark}

These operations are derived from the general principles of Archimedean
t-norm and t-conorms in \cite{beliakov2011}. The proposition
below follows directly from the observations stated in Remark \ref{remark17}.

\begin{proposition}
Consider two\ CC$q$-ROFVs, $A_{1}$ and $A_{2}$ and let $\lambda \geq0.$ The
operations $A_{1}\oplus_{q}A_{2}$, $A_{1}\otimes_{q}A_{2}$, $\lambda_{q}A_{1}%
$, $A_{1}^{\lambda_{q}}$ all yield results that are also CC$q$-ROFVs.
\end{proposition}

Aggregation operators play a crucial role in transforming multiple fuzzy input
values into a single representative output (see, \cite{beliakov2007, grabisch2009, klement2002}). We are now prepared to introduce two
aggregation operators; namely, the Gaussian weighted arithmetic aggregation
operator and the Gaussian weighted geometric aggregation operator; both
constructed for the collection of CC$q$-ROFVs.

\begin{definition}
Let $\left \{  A_{i}=\left \langle \mu_{A_{i}},\nu_{A_{i}},r_{A_{i}%
}\right \rangle :i=1,...,n\right \}  $ be a collection of CC$q$-ROFVs and
$Z:\left[  0,1\right]  \rightarrow \left[  0,\infty \right]  $ be the additive
generator of a continuous Archimedean t-norm or t-conorm. Then the
CC$q$-ROFWA aggregation operator is defined by
\[
\text{CC}q\text{-ROFWA}\left(  A_{1},...,A_{n}\right)  =\left(  Z\right)
\bigoplus \limits_{i=1}^{n}\omega_{i}A_{i},
\]
where $0\leq \omega_{i}\leq1$ for any $i=1,...,n$ with $\displaystyle\sum \limits_{i=1}%
^{n}\omega_{i}=1.$
\end{definition}

The CC$q$-ROFWA can be expressed as follows.

\begin{theorem}
Let $\left \{  A_{i}=\left \langle \mu_{A_{i}},\nu_{A_{i}},r_{A_{i}%
}\right \rangle :i=1,...,n\right \}  $ be a collection of CC$q$-ROFVs and let
$Z$ be the additive generator of the Algebraic t-norm, namely, $Z(t)=-\log t$.
Then the aggregation operator CC$q$-ROFWA is expressed as%
\[%
\begin{array}
[c]{lll}%
\text{CC}q\text{-ROFWA}\left(  A_{1},...,A_{n}\right)   & = & \left(  \left(  1-\left(
\operatorname{erf}^{-1}\left(  \prod \limits_{i=1}^{n}\frac{\operatorname{erf}%
\left(  1-\mu_{i}^{\operatorname{Re}}\right)  ^{\omega_{i}}}%
{\operatorname{erf}\left(  1\right)  ^{\omega_{i}-\frac{1}{n}}}\right)
\right)  ^{q}\right)  ^{\frac{1}{q}},\right.  \\
&  & \left.  \left(  1-\left(  \operatorname{erf}^{-1}\left(  \prod
\limits_{i=1}^{n}\frac{\operatorname{erf}\left(  1-\mu_{i}^{\operatorname{Im}%
}\right)  ^{\omega_{i}}}{\operatorname{erf}\left(  1\right)  ^{\omega
_{i}-\frac{1}{n}}}\right)  \right)  ^{q}\right)  ^{\frac{1}{q}}\right)  ,\\
&  & \left(  \operatorname{erf}^{-1}\left(  \prod \limits_{i=1}^{n}%
\frac{\operatorname{erf}\left(  \nu_{i}^{\operatorname{Re}}\right)
^{\omega_{i}}}{\operatorname{erf}\left(  1\right)  ^{\omega_{i}-\frac{1}{n}}%
}\right)  ,\operatorname{erf}^{-1}\left(  \prod \limits_{i=1}^{n}%
\frac{\operatorname{erf}\left(  \nu_{i}^{\operatorname{Im}}\right)
^{\omega_{i}}}{\operatorname{erf}\left(  1\right)  ^{\omega_{i}-\frac{1}{n}}%
}\right)  \right)  ,\\
&  & \left.  \prod \limits_{i=1}^{n}r_{A_{i}}^{\omega_{i}}\right \rangle .
\end{array}
\]

\end{theorem}

\begin{definition}
Let $\left \{  A_{i}=\left \langle \mu_{A_{i}},\nu_{A_{i}},r_{A_{i}%
}\right \rangle :i=1,...,n\right \}  $ be a collection of CC$q$-ROFVs and
$Z:\left[  0,1\right]  \rightarrow \left[  0,\infty \right]  $ be the additive
generator of a continuous Archimedean t-norm or t-conorm. The a CC$q$-ROFWG
aggregation operator is defined by
\[
\text{CC}q\text{-ROFWA}\left(  A_{1},...,A_{n}\right)  =\left(  z\right)
\bigotimes \limits_{i=1}^{n}A_{i}^{\omega_{i}}.
\]
\end{definition}

The CC$q$-ROFWG can be formulated as follows.

\begin{theorem}
Let $\left \{  A_{i}=\left \langle \mu_{A_{i}},\nu_{A_{i}},r_{A_{i}%
}\right \rangle :i=1,...,n\right \}  $ be a collection of CC$q$-ROFVs and let
$Z$ be the additive generator of the Algebraic t-norm, namely, $Z(t)=-\log t$. The
aggregation operator CC$q$-ROFWG is expressed as
\[%
\begin{array}
[c]{lll}%
\text{CC}q\text{-ROFWG}\left(  A_{1},...,A_{n}\right)   & = & \left \langle \left(
\operatorname{erf}^{-1}\left(  \prod \limits_{i=1}^{n}\frac{\operatorname{erf}%
\left(  \mu_{i}^{\operatorname{Re}}\right)  ^{\omega_{i}}}{\operatorname{erf}%
\left(  1\right)  ^{\omega_{i}-\frac{1}{n}}}\right)  ,\operatorname{erf}%
^{-1}\left(  \prod \limits_{i=1}^{n}\frac{\operatorname{erf}\left(  \mu
_{i}^{\operatorname{Im}}\right)  ^{\omega_{i}}}{\operatorname{erf}\left(
1\right)  ^{\omega_{i}-\frac{1}{n}}}\right)  \right)  ,\right.  \medskip \\
&  & \left(  \left(  1-\left(  \operatorname{erf}^{-1}\left(  \prod
\limits_{i=1}^{n}\frac{\operatorname{erf}\left(  1-\nu_{i}^{\operatorname{Re}%
}\right)  ^{\omega_{i}}}{\operatorname{erf}\left(  1\right)  ^{\omega
_{i}-\frac{1}{n}}}\right)  \right)  ^{q}\right)  ^{\frac{1}{q}}\right.
,\medskip \\
&  & \left.  \left(  1-\left(  \operatorname{erf}^{-1}\left(  \prod
\limits_{i=1}^{n}\frac{\operatorname{erf}\left(  1-\nu_{i}^{\operatorname{Im}%
}\right)  ^{\omega_{i}}}{\operatorname{erf}\left(  1\right)  ^{\omega
_{i}-\frac{1}{n}}}\right)  \right)  ^{q}\right)  ^{\frac{1}{q}}\right)  ,\\
&  & \left.  \prod \limits_{i=1}^{n}r_{A_{i}}^{\omega_{i}}\right \rangle
.\medskip
\end{array}
\]

\end{theorem}

\end{document}